\documentclass[12pt]{article}
\def\beq{\begin{equation}}
\def\eeq{\end{equation}}
\def\bea{\begin{eqnarray}}
\def\eea{\end{eqnarray}}
\def\bit{\begin{itemize}}
\def\eit{\end{itemize}}
\def\pa{\partial}

\def\de{\delta}

\def\Om{\Omega}
\def\al{\alpha}

\def\la{\lambda}

\def\eps{\epsilon}

\def\rd{{\rm d}}
\def\ri{{\rm i}}
\def\bJ{{\bf J}}

\def\Tr{{\rm Tr}}

\def\lra{\leftrightarrow}
\def\ra{\rightarrow}

\def\nn{\nonumber}

\def\tq{{\hat q}}
\def\tla{{\tilde \la}}

\begin{document}
\begin{center}

\vspace{1cm}
{\LARGE Eigenvalues of the volume operator\\ in loop quantum 
gravity}

\vspace{1cm}
{\large Krzysztof A. Meissner\footnote{E-mail address: {\tt 
Krzysztof.Meissner@fuw.edu.pl}}}

\vspace{0.5cm}
{\normalsize Institute of Theoretical Physics,
Warsaw University\\
\normalsize Ho\.za 69, 00-681 Warsaw, Poland\\
\vspace{0.2cm}
and\\
\vspace{0.2cm} 
CERN, Theoretical Division}

\end{center}

\begin{abstract}
We present a simple method to calculate certain sums of the eigenvalues 
of  the volume operator in loop quantum
gravity. We derive the asymptotic distribution of the eigenvalues in the
classical limit of very large spins which turns out to be of 
a very simple form. The results can be
useful for example in the statistical approach to quantum gravity.
\end{abstract}

\section{Introduction}

Volume operators were introduced by Rovelli and Smolin
\cite{rs} and by Ashtekar and Lewandowski
\cite{al} (the latter will be used in the current paper). The volume 
operator is much less understood than the area
operator, one of the reasons being that it is not diagonal in
any reasonable basis and there are ambiguities in defining its
action on a given graph (see \cite{alrep} and references therein; 
clarification of the issue is given in \cite{thgis}). The 
matrix elements of the volume operator were obtained in 
\cite{thiem}, \cite{depietri} (in the context of cosmology \cite{boj}); 
however, the formulae are very complicated and it is extremely difficult 
to get analytic expressions for the eigenvalues of the volume operator, 
except for very small values of the momenta (for some bounds on the 
eigenvalues see \cite{major}). Therefore it seems worthwhile 
to try to find some alternative way of describing the eigenvalues. In
this paper we show that relatively simple formulae can be obtained for
the sums of squares (or higher even powers) of the eigenvalues, with the 
most important ``quantum numbers'' in the vertex fixed: $j,j_1,j_2,j_3$, 
where $j$ is the total spin and $j_1,j_2,j_3$ are spins of the incoming 
legs. We also calculate the sums of higher powers (4, 6 and 8) of the 
eigenvalues with fixed  $j_1,j_2,j_3$. At the end
of the paper we derive (in the form of an integral) the distribution of 
the eigenvalues for fixed $j_1,j_2,j_3,j$ in the classical
limit, when all the values of spins tend to infinity, and show 
that without fixing $j$ the integral can be performed; the 
result turns out to be
surprisingly simple: $\rho(\la)=\arccos(\la/(j_1j_2j_3)/(2j_1j_2j_3)$ 
for $|\la|\le
j_1j_2j_3$ and $\rho(\la)=0$ otherwise.

\section{Volume operator}

The volume operator $V$ is defined as
\beq
V=\sqrt{|\tq|},
\eeq
where $\tq$ is analogous to the determinant of the metric
\beq
\tq=\sum_{I<J<K}\eps_{ijk} J_I^i\,J_J^j\, J_K^k.
\label{volop}
\eeq
Here, $I,J,K$ run through $1,2,...,N$ ($N\ge 3$) and $i,j,k$ are $SU(2)$
indices and we omit an overall constant proportional to the 
Planck volume. Eigenvalue of $V$ describes a contribution to the volume 
from a given vertex with $N+1$ legs with ``spin conservation'. We will 
concentrate on the most important 
case $N=3$ (so called 4-valent vertex) and later we will comment on the
application of the method to higher $N$.

The operator $\tq$ has several important properties:
\bit
\item
It commutes with all $\bJ_I^2$
\beq
[\tq,\bJ^2_I]=0.\ \ \ I=1,2,3
\label{VcomJ}
\eeq
so we can treat in the following all $\bJ_I^2=j_I(j_I+1)$ as fixed.

\item
It commutes with all components of the total spin $\bJ$
\beq
[\tq,\bJ]=0.\ \ \ \bJ:=\bJ_1+\bJ_2+\bJ_3
\label{tqbj}
\eeq
so its eigenvalues can be further indexed by the eigenvalues of $\bJ^2$,
denoted by $j(j+1)$. Note that because of (\ref{tqbj}) the eigenvalues of 
$\tq^2$ are the same for all $m$ in the $(j,m)$ multiplet. By this 
procedure, we are automatically in the gauge invariant case (where 
$\bJ_4:=-\bJ$ satisfies $\bJ_1+\bJ_2+\bJ_3+\bJ_4=0$).

\item
operator $\tq$ treats all four spins 
$\bJ_1,\bJ_2,\bJ_3,\bJ_4$ satisfying $\bJ_1+\bJ_2+\bJ_3+\bJ_4=0$ in the 
same way (modulo sign) since for example
\beq
\tq=\eps_{ijk} J_1^i\,J_2^j\, J_3^k=-\eps_{ijk} J_1^i\,J_2^j\, J_4^k
\eeq

\item
The operator $\tq$ therefore suggests a natural way to number states 
in a vertex: we specify $j_1,\ j_2,\ j_3,\ j,\ m$ of incoming legs and 
the eigenvalue of $\tq$ (the numbering is usually done in a 
``non--democratic'' way: spins $j_1$ and $j_2$ are composed to spin $j$, 
spins $j_3$ and $j_4$ are composed also to spin $j$ and the two spins $j$ 
are composed to a singlet)

\eit

Since the operator $\tq$ is self adjoint and (as can be easily seen from 
antisymmetry of $\eps_{ijk}$ and cyclic property of trace) trace 
of any odd power of $\tq$ vanishes, therefore $\tq$ has only real 
eigenvalues that are either 0 or come in pairs $(\la_i,-\la_i)$. We 
assume everywhere that all $j_I$ are bigger than 0 since if any of $j_I$ 
is equal to 0 then all the eigenvalues of the volume operator are also 
equal to 0.

As was shown in \cite{thiem},\cite{depietri} operator $\tq$ has very 
complicated
matrix elements as a function of $j,j_1,j_2,j_3$ (the matrix elements were 
obtained from 
the formula $\tq=\ri[\bJ_1\cdot\bJ_2,\bJ_2\cdot\bJ_3]$ and projection on a 
state with given $j$).

Note that there exists a convenient representation of the 
operator $\tq$  in the space of polynomials of degree $(2j_1+1)$ in $x$,  
$(2j_2+1)$ in $y$ and $(2j_3+1)$ in $z$ where
\beq
|j_1,m_1\rangle=\sqrt{(2j_1)!/(j_1-m_1)!/(j_1+m_1)!}\ x^{j_1+m_1}
\label{reprjm}
\eeq
and the generators
\beq
J_1^+ =-x^2\pa_x+2j_1 x,\ \ \ \ 
J_1^- =\pa_x,\ \ \ \ 
J_1^3=x\pa_x-j_1
\label{reprgen}
\eeq
and analogously for two other sets of generators. Then $\tq$ has the 
form:
\bea
\tq&=&\frac{\ri}{2}\left[-(x-y)(y-z)(z-x)\pa_x\pa_y\pa_z
+j_1(y-z)(y+z-2x)\pa_y\pa_z\right.\nn\\
&&+j_2(z-x)(z+x-2y)\pa_z\pa_x
+j_3(x-y)(x+y-2z)\pa_x\pa_y\nn\\
&&\left.+2j_1j_2(x-y)\pa_z+2j_2j_3(y-z)\pa_x+2j_3j_1(z-x)\pa_y\right]
\eea
As we can see explicitly, the $\tq$ operator conserves the total degree of 
the 
state (i.e. $m=m_1+m_2+m_3$) and indeed has complicated matrix elements 
especially after projection on eigenstates of $\bJ^2$.

\section{The square of the determinant operator} 

We consider the second power of the operator $\tq$:
\beq
\tq^2=\left(\eps_{ijk} J_1^i\,J_2^j\, J_3^k\right)^2.
\eeq
Using the well known formula
\bea
\eps_{ijk} \eps_{lmn} &=& \de_{il}\de_{jm}\de_{kn}-
\de_{jl}\de_{im}\de_{kn}+\de_{kl}\de_{im}\de_{jn}\nn\\
&&-\de_{il}\de_{km}\de_{jn}+\de_{jl}\de_{km}\de_{in}-
\de_{kl}\de_{jm}\de_{in}
\eea
we obtain
\bea
\tq^2&=&\bJ_1^2 \bJ_2^2 \bJ_3^2 -\bJ_1^2 J_2^i (\bJ_2\cdot \bJ_3)
J_3^i-\bJ_3^2 J_1^i (\bJ_1\cdot \bJ_2) J_2^i
-\bJ_2^2 J_1^i (\bJ_1\cdot \bJ_3) J_3^i\nn\\
&&+J_1^i (\bJ_1\cdot \bJ_2)(\bJ_2\cdot \bJ_3) J_3^i
+J_1^i J_2^j (\bJ_1\cdot \bJ_3)J_2^i J_3^j.
\label{V2JJ}
\eea

We want to trace the operator with the states
\beq
|(j_1,m_1)\otimes(j_2,m_2)\otimes(j_3,m_3)\rangle
\eeq
(often written as $|m_1,m_2,m_3\rangle$ for short), not yet
imposing any restriction on $j$ and $m=m_1+m_2+m_3$. The
calculation uses well known formulae (straightforward from (\ref{reprjm}) 
and (\ref{reprgen})):
\bea
J_I^+ |j_I,\,m_I\rangle&=&\sqrt{j_I(j_I+1)-m_I(m_I+1)}\ 
|j_I,\,m_I+1\rangle \nn\\
J_I^- |j_I,\,m_I\rangle&=&\sqrt{j_I(j_I+1)-m_I(m_I-1)}\ 
|j_I,\,m_I-1\rangle \nn\\
J_I^3 |j_I,\,m_I\rangle&=&m_I\ 
|j_I,\,m_I\rangle 
\label{jpjm}
\eea
The result reads
\bea
&&\langle m_1,m_2,m_3|\,\tq^2\,
|m_1,m_2,m_3\rangle=\nn\\
&&\ \ \ \ \ \frac12 m_1^2\left[\left(j_2(j_2+1)-m_2^2\right)
\left(j_3(j_3+1)-m_3^2\right)-m_2 m_3\right]\nn\\
&&
\ \ \ \ \ +\frac12 m_2^2\left[\left(j_1(j_1+1)-m_1^2\right)
\left(j_3(j_3+1)-m_3^2\right)-m_1 m_3\right]\nn\\
&&\ \ \ \ \ +\frac12 m_3^2\left[\left(j_1(j_1+1)-m_1^2\right)
\left(j_2(j_2+1)-m_2^2\right)-m_1 m_2\right].
\label{TrV2}
\eea

Summing over all $m_1,m_2,m_3$, we have the result
\beq
\Tr\, \tq^2=\frac29 j_1(j_1+1)(2j_1+1)j_2(j_2+1)(2j_2+1)j_3(j_3+1)
(2j_3+1).
\label{trallV}
\eeq
Since the total number of states is equal to
$(2j_1+1)(2j_2+1)(2j_3+1)$ the average eigenvalue of $\tq^2$ is
equal to
\beq
\langle \tq^2\rangle=\frac29 j_1(j_1+1)j_2(j_2+1)j_3(j_3+1).
\eeq
In the last section we will derive the general formula for $\langle 
\tq^{2N}\rangle$ in the limit of large $j$'s.

\section{Sum of the eigenvalues with fixed $j$}

In this section we calculate $\Tr\ \tq^2$ with an additional condition
\beq
m_1+m_2+m_3=m,
\label{mcond}
\eeq
where to be definite we take $m\le 0$.
We sum in this way many different $j$ multiplets with $j\ge -m$; it is
important to notice that because of (\ref{VcomJ}) the eigenvalues of 
$\tq^2$ are the same for all $(j,m)$ components of a given $j$ multiplet.

As one can easily convince oneself it is a non-trivial problem to
impose the condition (\ref{mcond}) directly. Therefore we use an indirect 
method that will partly do the bookkeeping for us: we perform the sum 
with the additional factor
\beq
f_2(w,j_1,j_2,j_3):=\Tr\,\left( \tq^2 w^{J_1^3+J_2^3+J_3^3}\right)
\label{TrV2w}
\eeq
and at the end we will get to the final result by picking up the 
coefficient in front of $w^m$.

We use the formula
\beq
\sum_{m_1=-j_1}^{j_1} m_1^p w^{m_1}=(w\pa_w)^p\,\frac{w^{-j_1}-w^{j_1+1}}
{1-w}.
\eeq

The computation of $f_2(w,j_1,j_2,j_3)$ from (\ref{TrV2}) gives
\bea
&&(1-w)f_2(w,j_1,j_2,j_3)=\frac{2}{(1-w)^8}\,
w^{-j_1-j_2-j_3+1}\left\{j_1j_2j_3(j_1+j_2+j_3)\right.\nn\\
&&-w\left[j_1j_2((j_1+j_2)(2j_3+1)+2j_3-1)+(1\lra 3)+(2\lra
3)\right]\nn\\
&&+w^2\left[(j_1+1)(j_1+\frac32 (j_2+j_3)(2j_1-1)+6j_1j_2j_3)+(1\lra 2)
+(1\lra 3)\right]\nn\\
&&-w^3\left[(j_1+1)(j_2+1)((j_1+j_2)(2j_3+1)+2j_3-1)+(1\lra 3)+(2\lra
3)\right]\nn\\
&&\left.+w^4(j_1+1)(j_2+1)(j_3+1)(j_1+j_2+j_3+3)\right\}\nn\\
&&-(j_1\ra -j_1-1)-(j_2\ra -j_2-1)-(j_3\ra -j_3-1)\nn\\
&&+(j_1\ra -j_1-1,j_2\ra -j_2-1)+(j_2\ra -j_2-1,j_3\ra -j_3-1)\nn\\
&&+(j_1\ra -j_1-1,j_3\ra -j_3-1)\nn\\
&&-(j_1\ra -j_1-1,j_2\ra -j_2-1,j_3\ra -j_3-1).
\label{fw}
\eea
To obtain the sum of squares of eigenvalues with fixed $j$ we subtract 
the result for $f_2(w,j_1,j_2,j_3)$ with $m=-j$ from the result
for $m=-j-1$ (it follows from the fact that the eigenvalues are the same 
for all $m$ in $(j,m)$ multiplets). We therefore expand
$(1-w)f_2(w,j_1,j_2,j_3)$ given in (\ref{fw}) in $w$, and the coefficient 
in front of $w^{-j}$ is the desired sum of squares of eigenvalues with 
fixed $j$. Hence for example:
\begin{itemize}
\item
the 
eigenvalue for 
$j=(j_1+j_2+j_3)$ is equal to 0
\item
for
$j=(j_1+j_2+j_3-1)$
eigenvalues are equal to $\pm\sqrt{j_1 j_2 j_3 (j_1+j_2+j_3)}$
\item
for 
$j=(j_1+j_2+j_3-2)$ there is an eigenvalue 0 and two eigenvalues equal to
$\pm\sqrt{(4j_1 j_2 j_3-j_1 j_2-j_2 j_3-j_1 j_3) (j_1+j_2+j_3-1)+j_1 j_2 
j_3}$ (if any of the $j_I$ is equal to 1/2 then some of these 
eigenvalues may be missing).
\end{itemize}

We may note that for the $n$-valent vertex ($n>4$) the method works 
in exactly the same way: in the $\tq^2$ operator, 
beyond the already discussed terms, we encounter also terms of the form
\beq
\eps_{ijk}J_1^i J_2^j J_3^k \eps_{lmn}J_1^l J_2^m J_4^n.
\eeq
After sandwiching between $ |m_1,m_2,m_3,m_4\rangle$ we have to
assign $k=3$, $n=3$ and therefore $(i,j,k,l)=(+,-,-,+)$ or
$(i,j,k,l)=(-,+,+,-)$, so that the appropriate term in
$f_2(w,j_1,j_2,j_3,j_4)$ reads
\beq
\sum_{m_1,m_2,m_3,m_4} w^{m_1+m_2+m_3+m_4}
 \frac12 m_3 m_4\left(j_1(j_1+1)-m_1^2\right)
\left(j_2(j_2+1)-m_2^2\right)
\eeq
and the sum can be performed without difficulty.

As can be easily seen the terms of the form $\eps_{ijk}J_1^i J_2^j J_3^k 
\eps_{lmn}J_1^l J_4^m 
J_5^n$ or $\eps_{ijk}J_1^i J_2^j J_3^k \eps_{lmn}J_4^l J_5^m J_6^n$
give no contribution to $f_2$.

We have also calculated along the same lines 
\beq  
f_4(w,j_1,j_2,j_3):=\Tr\,\left( \tq^4 w^{J_1^3+J_2^3+J_3^3}\right)
\label{TrV4w}
\eeq
but the resulting formulae are too long to be presented here.

\section{The density distribution for eigenvalues}

It would be very desirable to have also higher-order moments,
\beq
\Tr\ \tq^{2n}|_{j,j_1,j_2,j_3},
\eeq
since we could then reconstruct the full density by inverse Mellin
transform of the relation
\beq
\Tr\ \tq^{2n}|_{j,j_1,j_2,j_3}=\int\rd\la \rho_j(\la)\la^{2n}.
\eeq
Since $\tq$ is an operator in a finite-dimensional space, the density 
distribution is just a sum of Dirac delta functions; we aim at finding a 
smooth approximation of this distribution for large values of the spins 
involved.

In this section we will calculate higher moments
($\Tr\ \tq^4$, $\Tr\ \tq^6$ and $\Tr\ \tq^8$)  as a function of $j_1$, 
$j_2$, and $j_3$, but with $j$ not fixed.

Introducing the notation
\beq
\langle B\rangle:=\frac{\Tr B}{\Tr {\bf 1}},
\eeq
we have the relation
\bea
\langle e^{-A}\rangle&=&\exp\left(-\langle A\rangle+\frac12(\langle
A^2\rangle-\langle A\rangle^2)\right.\nn\\
&&-\frac16(\langle A^3\rangle-3\langle A^2\rangle\langle
A\rangle+2\langle
A\rangle^3)\\
&&\left.+\frac1{24}(\langle A^4\rangle-4\langle A^3\rangle\langle
A\rangle-3\langle A^2\rangle^2+12\langle A^2\rangle\langle
A\rangle^2-6\langle A\rangle^4)+\ldots\right).\nn
\eea
Neglecting terms of the third order and higher in the exponent (Gaussian
approximation) we have
\beq
\langle e^{-\al\tq}\rangle\approx
e^{-\langle \al\tq\rangle+\frac{\al^2}{2}(\langle
\tq^2\rangle-\langle \tq\rangle^2)}=\int\rd\la \rho_G(\la) e^{-\al\la}
\eeq
and since $\langle \tq\rangle=0$ we have
\beq
\rho_G(\la)=\sqrt{\frac{1}{2\pi\beta}}e^{-\la^2/(2\beta)},
\eeq
where
\beq
\beta=\langle \tq^2\rangle.
\eeq
In the Gaussian approximation
\beq
\langle \tq^{2n}\rangle=(2n-1)!!\,\langle \tq^2\rangle^n.
\eeq

As we already know,
\beq
\langle \tq^2\rangle= \frac29 F_3,
\label{V2}
\eeq
where we introduced
\bea
F_0&=&1,\nn\\
F_1&=&j_1(j_1+1)+ j_2(j_2+1)+ j_3(j_3+1),\nn\\
F_2&=&j_1(j_1+1) j_2(j_2+1) +j_1(j_1+1)
j_3(j_3+1)+ j_2(j_2+1) j_3(j_3+1),\nn\\
F_3&=&j_1(j_1+1) j_2(j_2+1) j_3(j_3+1).
\eea

Using the form (\ref{V2JJ}) we obtain (with the help of the MAPLE program)
\beq
\langle \tq^4\rangle=\frac{F_3}{225}
\left(24 F_3-8 F_2+11 F_1-12 F_0\right),
\label{V4}
\eeq
\bea
\langle \tq^{6}\rangle&=&
\frac{F_3}{22050}
\left(1440 F_3^2-1440 F_3 F_2+480  F_2^2+2538 F_3 F_1
-1656 F_2 F_1\right.\nn\\
&&\left.\ \ \ -6909 F_3 F_0+1042 F_1^2+5158 F_2 F_0-4815 F_1 F_0+3222 
F_0^2\right)
\label{V6}
\eea
and
\bea
\langle \tq^{8}\rangle&=&
\frac{F_3}{99225}\left(
4480 F_3^3 - 8960 F_3^2 F_2 + 8064 F_3 F_2^2 + 19936 F_3^2 F_1\right.\nn\\ 
&&- 78032 F_3^2 F_0- 2688 F_2^3 - 38304 F_3 F_2 F_1+180204 F_3 F_2 \nn\\ 
&&+ 41853 F_3 F_1^2 + 18480 F_2^2 F_1 -385701 F_3 F_1 F_0 - 91932 
F_2^2 F_0\nn\\
&& - 36351 F_2 F_1^2
+691740 F_3 F_0^2 + 329841 F_2 F_1 F_0 + 18081 F_1^3\label{V8}\\
&&\left.-549288 F_2 F_0^2-204552 F_1^2 F_0+521316 F_1 F_0^2-285282 
F_0^3\right).\nn
\eea
The result for $\langle \tq^{2n}\rangle$ has the form $F_3\cdot P_{n-1}$ 
where $P_{n-1}$ is a polynomial with alternating signs, 
homogeneous of degree $(n-1)$ in $F_3,\ F_2,\ F_1,\ F_0$.

We were unable to find a general expression for the coefficients of these 
polynomials, however, as will be shown in the next section, we were able 
to prove that  for $\langle\tq^{2n}\rangle$ the coefficient 
in front of $F_3^n$, which is the most important term for large spins, is 
equal to $\frac{(2n)!!}{(2n+1)(2n+1)!!}$ and for $n>1$ it differs from the 
Gaussian approximation result $(2n-1)!!(2/9)^n$.

\section{Classical distribution of the eigenvalues}

In this section we will find the classical distribution (both integrated 
$\rho(\la)$ and in the subspace of fixed total spin $\rho_j(\la)$)of 
the eigenvalues of the operator $\tq$ in the limit of large spins. In this 
limit we can disregard any commutators of $\bJ_I$ with itself (since they 
introduce lower power of spins) and the case boils down 
to the classical density. 

For the classical density we have,
\bea
\int_{-j_1j_2j_3}^{j_1j_2j_3} \rd\la\,\rho_j(\la)\, \la^{2n}&=&
N^{-1}\int \frac{\rd\Om_1 \rd\Om_2 \rd\Om_3}{(4\pi)^3}
\left((\bJ_1\times \bJ_2)\cdot\bJ_3 \right)^{2n}\cdot\nn\\
&&
\de\left((\bJ_1+\bJ_2+\bJ_3)^2-j^2\right),
\eea
where we integrate over angular positions of $\bJ_1$, $\bJ_2$ 
and $\bJ_3$, and $N$ is a normalization factor:
\beq
N=\int \frac{\rd\Om_1 \rd\Om_2 \rd\Om_3}{(4\pi)^3}
\de\left((\bJ_1+\bJ_2+\bJ_3)^2-j^2\right).
\eeq

We introduce also the integrated density satisfying
\beq
\int_{-j_1j_2j_3}^{j_1j_2j_3} \rd\la\,\rho(\la)\, \la^{2n}=
\int \frac{\rd\Om_1 \rd\Om_2 \rd\Om_3}{(4\pi)^3}\left((\bJ_1\times 
\bJ_2)\cdot\bJ_3
\right)^{2n}.
\eeq

Since only relative angles are important, we can fix some positions:
\beq
\theta_3=0,\ \ \ \phi_2=0
\eeq
and then for example
\beq
(\bJ_1\times \bJ_2)\cdot\bJ_3=j_1 j_2 j_3\sin\theta_1 \sin\theta_2 
\sin\phi.
\eeq

We introduce coordinates on $S^3$:
\bea
x&=&\sin\theta_1 \sin\theta_2 \sin\phi,\nn\\
y&=&\sin\theta_1 \sin\theta_2 \cos\phi,\nn\\
z&=&\sin\theta_1 \cos\theta_2, \nn\\
t&=&\cos\theta_1.
\eea

Then
\bea
\int_{-j_1j_2j_3}^{j_1j_2j_3} \rd\la\,\rho_j(\la)\, \la^{2n}&=&
\frac1{8\pi N}\int \frac{\rd x\rd y\rd z \rd 
t}{\sqrt{1-t^2}}\,2\de\left(x^2+y^2+z^2+t^2-1\right)\nn\\
&&\left(j_1 j_2 j_3 x \right)^{2n}\de\left(j_1^2+j_2^2+j_3^2+2j_1j_2 t+
2j_1j_3\frac{z}{\sqrt{1-t^2}}\right.\nn\\
&&\left.\ \ \ \ \ \ 
+2j_2j_3\frac{zt}{\sqrt{1-t^2}}+2j_2j_3 y-j^2\right).
\eea

Comparing the sides we obtain the final result, i.e. the 
classical distribution function of eigenvalues in terms of 
$j_1,\ j_2,\ j_3,\ j$: 
\bea
\rho_j(\la)&=&
\frac{1}{8\pi Nj_1j_2j_3}\int \frac{\rd y\rd z \rd
t}{\sqrt{1-t^2}}2\de\left(\tla^2+
y^2+z^2+t^2-1\right)\nn\\
&& 
\de\left(j_1^2+j_2^2+j_3^2+2j_1j_2 t+
2j_1j_3\frac{z}{\sqrt{1-t^2}}\right.\nn\\
&&\left.\ \ \ +2j_2j_3\frac{zt}{\sqrt{1-t^2}}+2j_2j_3 y-j^2\right),
\eea
where 
\beq
\tla=\la/(j_1j_2j_3).
\eeq

We can integrate over $t$: 
\bea
\rho_j(\la)&=&
\frac{1}{8\pi Nj_1j_2j_3}\int_0^{\sqrt{1-\tla^2}}\int_0^{2\pi} 
\frac{r\rd 
r\rd\phi}{\sqrt{r^2+\tla^2}\sqrt{1-r^2-\tla^2}}\nn\\
&&
\sum_{\epsilon=\pm 1}\de\left(j_1^2+j_2^2+j_3^2+2j_1j_2\epsilon 
\sqrt{1-r^2-\tla^2}+  
2j_1j_3\frac{r\sin\phi}{\sqrt{r^2+\tla^2}}\right.\nn\\
&&\left.\ \ \ 
+2j_2j_3\epsilon 
\frac{r\sin\phi\sqrt{1-r^2-\tla^2}}{\sqrt{r^2+\tla^2}}+2j_2j_3 
r\cos\phi-j^2\right).
\eea

It is difficult to perform the integral in this form. However, performing 
the same steps for the integrated distribution function we obtain a 
surprisingly simple result:
\beq
\rho(\la)=  
\frac{1}{2j_1j_2j_3}\int_0^{\sqrt{1-\tla^2}}
\frac{r\rd r}{\sqrt{r^2+\tla^2}\sqrt{1-r^2-\tla^2}}
=\frac{1}{2j_1j_2j_3}
\arccos\,(\tla)
\eeq
for $|\tla|\le 1$ and 0 for $|\tla|>1$.

Therefore we get the coefficient in front of $F_3^n$ for the integrated 
trace
\beq
\langle \tq^{2n}\rangle \to
\int_{-1}^1\rd x \frac12\arccos(x) 
x^{2n}F_3^{n}=\frac{(2n)!!}{(2n+1)(2n+1)!!}F_3^{n}.
\eeq
Hence for example we recover the leading coefficients of 
(\ref{V2}), (\ref{V4}),  (\ref{V6}) and  (\ref{V8}):
\beq
\langle \tq^2\rangle\to\frac29 F_3,\ \ \ \langle  \tq^4\rangle 
\to\frac8{75} F_3^2,\ \ \ \ 
\langle\tq^6\rangle\to\frac{16}{245} F_3^3,\ \ \ \ 
\langle\tq^8\rangle\to\frac{128}{2835} F_3^4.
\eeq

Concluding, by using the methods described in this paper we can learn 
something about 
the distribution of the eigenvalues of the volume operator in 
the quantum case, and there exists an explicit formula for this 
distribution in the limiting case of large spins, i.e. the classical 
case. It would
be interesting to compare these results with the numerical analysis for   
low spins (up to 50) given in \cite{thiem}.

Besides being mathematically interesting, these results can be important 
for example in the statistical approach to loop quantum gravity. 

Several conceptual problems arise in this context by the results 
of the present paper. The first one is what observables (besides the 
volume) should be specified to actually describe a 
macroscopic state in a given region. Since the contribution to the 
volume from 
a given 4-valent vertex grows like $j_1\cdot j_2\cdot j_3$ it is important 
to know whether these 
observables provide suppression of the contribution 
from very large spins rendering final expectation value of the volume 
finite. 
The second problem is connected with the first one and concerns the bulk 
entropy: there are many sets of 
graphs with many different assignments of spins on legs that give 
(macroscopically) the same  volume. If we specify both the volume and the 
other observables we are left with a certain number of graphs. The 
question is, 
whether the logarithm of this number is connected with the
bulk entropy much as sequences of spins are connected with 
the black hole surface entropy \cite{bhe}.

\vspace{0.5cm}

\noindent {\bf Acknowledgements}

The author would like to thank Hermann Nicolai and the whole group
at the AEI for hospitality. Discussions with Abhay Ashtekar, Johannes 
Brunnemann, Thomas 
Thiemann and especially Jurek Lewandowski are gratefully acknowledged. 
This work was partially supported by the Polish KBN
grant 2P03B 001 25 and the European Program HPRN-CT-2002-0277.


\begin{thebibliography}{Ref}

\bibitem{rs} C. Rovelli and L. Smolin, ``Discreteness of area and
volume in quantum gravity'', Nucl. Phys. {\bf B442} (1995) 593;
Erratum, Nucl. Phys. {\bf B456} (1995) 734.

\bibitem{al} A. Ashtekar and J. Lewandowski, ``Differential geometry on
the space of connections via graphs and projective limits'', J.
Geom.  Phys. {\bf 17} (1995) 191;\\
``Quantum theory of geometry, II: Volume operators'', Adv. Theor.
Math. Phys. {\bf 1} (1998) 388, {\tt gr-qc/9711031}.

\bibitem{alrep}
A.~Ashtekar and J.~Lewandowski,
``Background independent quantum gravity: A status report,''
Class.\ Quant.\ Grav.\  {\bf 21} (2004) R53,
{\tt gr-qc/0404018}.

\bibitem{thgis}  K. Giesel and T. Thiemann, 
Consistency Check on Volume and Triad Operator Quantisation in Loop 
Quantum Gravity,  {\tt gr-qc/0507036},  {\tt gr-qc/0507037},

\bibitem{thiem} T. Thiemann, ``Closed formula for the matrix elements of
the volume operator in  canonical quantum gravity'',
J. Math. Phys. {\bf 39} (1998) 3347, {\tt gr-qc/9606091},\\
J. Brunnemann and T. Thiemann, `Simplification of the spectral
analysis of the volume operator in loop quantum gravity'', {\tt
gr-qc/0405060}.

\bibitem{depietri}
R.~De Pietri and C.~Rovelli,
``Geometry eigenvalues and scalar product from recoupling 
theory in loop quantum gravity,''
Phys.\ Rev.\ D {\bf 54} (1996) 2664, {\tt gr-qc/9602023}.

\bibitem{boj}
M.~Bojowald,
``Loop quantum cosmology. II: Volume operators'',
Class.\ Quant.\ Grav.\  {\bf 17} (2000) 1509,
{\tt gr-qc/9910104}.


\bibitem{major}
S.~A.~Major and M.~D.~Seifert,
``Modelling space with an atom of quantum geometry,''
Class.\ Quant.\ Grav.\  {\bf 19}, 2211 (2002)
{\tt gr-qc/0109056}.

\bibitem{bhe}
A.~Ashtekar, J.~Baez, A.~Corichi and K.~Krasnov,
``Quantum geometry and black hole entropy'',
Phys.\ Rev.\ Lett.\  {\bf 80} (1998) 904, {\tt gr-qc/9710007}\\
M.~Domagala and J.~Lewandowski,
Class.\ Quant.\ Grav.\  {\bf 21} (2004) 5233,
{\tt gr-qc/0407051}\\
K.~A.~Meissner,
``Black hole entropy in loop quantum gravity,''
Class.\ Quant.\ Grav.\  {\bf 21} (2004) 5245, {\tt gr-qc/0407052}.

\end{thebibliography}
\end{document}